\def\rn{\noindent\parshape 2 0truecm 8.8truecm 0.3truecm 8.5truecm}
\def\nn#1 #2{#1, #2.}				% Name with 1 initial
\def\nnn#1 #2 #3{#1, #2. #3.}			% Name with 2 initials
\def\nnnn#1 #2 #3 #4{#1, #2. #3. #4.}		% Name with 3 initials
\def\nnnnn#1 #2 #3 #4 #5{#1, #2. #3. #4. #5.}	% Name with 4 initials
\def\dualand{, \&\hbox{ }}				% Lower case "and" already in use.
\def\multiand{, \&\hbox{ }}				% Lower case "and" already in use.
\def\rg#1;#2;#3;#4;#5;#6 {\par\rn#1 #2, {\it #3}, {\bf #4}, #5 (``#6'') \par}
\def\rf#1;#2;#3;#4;#5 {\par\rn#1 #2, {\it #3}, {\bf #4}, #5\par}
\def\rfbook#1;#2;#3;#4;#5 {{\frenchspacing\par\rn#1 #2, {\it #3} (#4: #5)\par}}
\def\rfproc#1;#2;#3;#4;#5;#6 {{\frenchspacing\par\rn#1 #2, in {\it #3}, ed. #4 (#5: #6)\par}}
\def\rfprep#1;#2;#3  {{\par\rn#1 #2, #3\par}}
\def\rfprepp#1;#2;#3 {{\par\rn#1 #2, #3\par}}

\def\etal{{\frenchspacing\it et al.}}
\def\ie{{\frenchspacing\it i.e.}}
\def\eg{{\frenchspacing\it e.g.}}

\def\beq#1{\begin{equation}\label{#1}}
\def\eeq{\end{equation}}
\def\beqa#1{\begin{eqnarray}\label{#1}}
\def\eeqa{\end{eqnarray}}

\def\spose#1{\hbox to 0pt{#1\hss}}
\def\simlt{\mathrel{\spose{\lower 3pt\hbox{$\mathchar"218$}}
     \raise 2.0pt\hbox{$\mathchar"13C$}}}
\def\simgt{\mathrel{\spose{\lower 3pt\hbox{$\mathchar"218$}}
     \raise 2.0pt\hbox{$\mathchar"13E$}}}
\def\simpropto{\mathrel{\spose{\lower 3pt\hbox{$\mathchar"218$}}
     \raise 2.0pt\hbox{$\propto$}}}

\def\ed{\end{document}}

\def\Ob{\Omega_b}
\def\Oc{\Omega_{cdm}}
\def\Ok{\Omega_k}
\def\Ol{\Omega_\Lambda}
\def\Om{\Omega_m}

\def\ob{\omega_b}
\def\oc{\omega_{cdm}}

\def\om{\omega_{cdm}}

\def\Cl{C_\l}

\def\ns{n_s}
\def\nt{n_t}
\def\dA{d_A}
\def\zlss{z_{lss}}

\def\p{{\bf p}}

\def\l{\ell}
\def\Cl{C_\ell}
\def\Cl{C_\ell}

\documentstyle[emulateapj]{article}
\begin{document}

%%%%%%%%%%%%%%%%%%%%%%%%%%%%%

%\tighten
%\eqsecnum
%\received{4 August 1988}
%\accepted{23 September 1988}
\journalid{337}{15 January 1989}
\articleid{11}{14}

%\submitted{Submitted to ApJL September 16, 1998}
\submitted{Submitted to ApJL September 16; accepted February 2}
%\submitted{To be submitted to ApJL}

\title{Cosmological constraints from current CMB and SN 1a data:
a brute force 8 parameter analysis}

% \title{CMB constraints on 9 parameter CDM model}

\author{
Max Tegmark
\footnote{Institute for Advanced Study, Princeton, 
NJ 08540; max@ias.edu}$^,$\footnote{Hubble Fellow}
}

\begin{abstract}

We describe constraints on a ``standard'' 8 parameter 
open cold dark matter (CDM) model from the most recent 
CMB and SN1a data.
Our parameters are the densities of CDM, baryons, vacuum energy and
curvature, the reionization optical depth, and the
normalization and tilt for both scalar and tensor fluctuations.
We find that although the possibility of reionization
and gravity waves substantially weakens the 
constraints on CDM and baryon density, tilt, Hubble constant
and curvature, allowing {\eg} a closed Universe,
open models with vanishing cosmological constant are still strongly 
disfavored.
% We find that since the constraints are still weak, the choice of 
% statistical test makes a substantial difference. We present 
% Bayesian and prior-independent frequentist constraints.

\end{abstract}

%\preprint{IASSNS-AST 97/666}
%\date{\today}
%\date{Submitted November 21, 1996; accepted February 26, 1997}

%%%%%%%%%%%%%%%%%%%%%%%%%%%%%%%%%%%%%%%%%%%%%%
%%%%%%%%%%%%%%%%%%%%%%%%%%%%%%%%%%%%%%%%%%%%%%

\section{INTRODUCTION}

The currently most popular cosmological model has of order 
$N=10$ free parameters.
Upcoming CMB experiments hold the potential to 
measure these parameters with
unprecedented accuracy (Jungman {\etal} 1996; 
Bond {\etal} 1997; Zaldarriaga 
{\etal} 1997; Efstathiou \& Bond 1998),
especially when combined with galaxy redshift surveys
(Eisenstein {\etal} 1998) and supernovae 1a (SN 1a) 
observations (White 1998; Tegmark {\etal} 1998).
However, these papers have also demonstrated the importance
of fitting for all $N$ parameters jointly,
revealing subtle degeneracies by exploring the full 
$N$-dimensional parameter space.
% Although it is tempting to reduce $N$ by invoking theoretical prejudice
% for the values of some parameters (the $N=2$ case being particularly 
% desirable since it is easy to plot), 
% this tends to give misleadingly small error bars. 
For this reason, there has been a persistent drive towards
larger $N$ when analyzing data. The first analyses based on 
COBE DMR used $N=2$ parameters, the CMB quadrupole normalization 
$Q$ and the scalar tilt $\ns$ of the power spectrum
(\eg, Smoot {\etal} 1992; Gorski {\etal} 1994; Bond 1995;  
Bunn \& Sugiyama 1995; Tegmark \& Bunn 1995).
Since then, many dozens of papers have extended this
to incorporate more data and parameters, recent work including 
Bunn \& White (1997); de Bernardis {\etal} (1997); Ratra {\etal} (1998); Hancock
{\etal} (1998); Lesgourges {\etal} (1998); Bartlett {\etal} (1998); 
Webster {\etal} (1998); Lineweaver \& Barbosa (1998ab); 
White (1998); Bond \& Jaffe (1998); 
Gawiser \& Silk (1998), and
Contaldi {\etal} (1998). 
The most ambitious analysis to date is that of
Lineweaver (1998 -- hereafter L98), jointly varying
$N=6$ parameters:
$n_s$, $Q$, the Hubble constant $h$ and
the relative densities $\Oc$, $\Ob$ and $\Ol$
of CDM, baryons and vacuum energy.

A realistic minimal cosmological model should include all 
physically well-motivated parameters. Yet
even the heroic L98 analysis lacks three:
gravity-wave (tensor) fluctuations, parametrized by a 
relative quadrupole normalization $r$
and a tilt $\nt$, and the optical depth $\tau$ from reionization.
In an inflationary context, gravity waves are just as natural as 
deviations from $\ns=1$, and we know that $\tau>0$ since the Universe
was reionized before $z=5$. % (Gunn \& Petersen REF).
It is therefore timely to extend this drive towards larger $N$ by
analyzing this ``minimal'' 9-parameter model space.
That is the purpose of the present {\it Letter}. 
% Yet even this heroic
% analysis lacks three well-motivated parameters that should be included in
% a realistic minimal cosmological model:
% gravity-wave (tensor) fluctuations, parametrized by a 
% relative quadrupole normalization $r$
% and a tilt $\nt$, and the optical depth $\tau$ from reionization.
% In an inflationary context, gravity waves are just as natural as 
% deviations from $\ns=1$, and we know that $\tau>0$ since the Universe
% was reionized before $z=5$. % (Gunn \& Petersen REF).
% It is therefore timely to complete this drive towards larger $N$ by
% performing a full 8 parameter analysis. 
% This is the purpose of the present {\it Letter}. 

% The previous work most similar to this letter is the $N=6$ 
% analysis in L98. 

\section{METHOD}

In principle, such an analysis is straightforward: compute 
the theoretical CMB power spectrum $C_\l$
with the CMBfast software (Seljak \& Zaldarriaga 1996) 
at a fine grid of points in 
the $N$-dimensional parameter space and make $\chi^2$-fits
to the available power spectrum measurements in Figure 1.  
In practice, this is quite tedious. With $M$ grid points in each 
dimension, $M^N$ power spectra must be computed.
Lineweaver's impressive $N=6$ analysis involved running CMBfast 
millions of times, corresponding to years of workstation CPU time,
and with $M\sim 20$ as in L98, the amount of 
work grows by more than an order of magnitude for each additional parameter.
% Thus if one uses $M\sim 20$ as in L98, the amount of 
% work grows by more than an order of magnitude for each additional parameter.
% Fortunately, as described below, we can take advantage of the underlying physics
% to save substantial amounts of time.
Fortunately, the underlying physics
(see {\eg} Hu {\etal} 1997 for a review) 
allows 
%%% several 
numerical simplifications as described below.

\subsection{Parameter space}

We choose our 9-dimensional parameter vector to be
$\p\equiv(\om,\ob,\tau,h,\Ok,\ns,\nt,Q,r)$,
where the physical densities 
$\omega_i\equiv h^2\Omega_i$, $i=cdm,\>b$. The advantage of this 
parametrization (see Bond {\etal} 1997; Eisenstein {\etal} 1998)
will become clear in \S\ref{HighLowSec}.
$\Ok$ is the spatial curvature, so 
%in terms of these parameters, 
$\Ol=1-\Ok-\Oc-\Ob=1-\Ok-(\oc+\ob)/h^2$. 
% As discussed in {\eg} Bond {\etal} (1997) and
% Eisenstein {\etal} (1998), the quantities 
% $\omega_i\equiv h^2\Omega_i$ are better to work with than $\Omega_i$ 
% ($i=m,b$), since they correspond to the physical densities 
% occurring in the Boltzmann equation. 
% Since the curvature $\Ok\equiv 1-\Om-\Ol$, 
% varying $h$ with the other parameters fixed simply changes 
% $\Ol$, and has no other effect than shifting the power spectrum sideways
% and adding a late integrated Sachs-Wolfe effect at the very lowest $\l$.

We choose our grid to cover the following parameter ranges:
$0.02\le\om\le 0.8$, 
$0.003\le\ob\le 0.13$, 
$0\le\tau\le 0.8$, 
$0.2\le h\le 1.3$, 
$0\le\Ok\le 0.9$,
$0.5\le\ns\le 1.6$, 
$0.24\le\nt\le 1$. 
This extends the L98 ranges somewhat, since L98 reported high 
likelihoods near certain grid boundaries. 
To avoid prohibitively large $M$,
we use a roughly logarithmic
grid spacing for $\om$, $\ob$ and $h$, 
a linear grid spacing for $\Ok$, 
a hybrid for $\ns$, $\nt$, $\tau$ and no grid at all for the 
%%% multiplicative
normalization factors $Q$ and $r$.

% \subsection{Adaptive mesh and interpolation}

Although a fairly fine grid is desirable for the likelihood 
analysis presented in Section 3, we find that we can attain sufficient 
accuracy by running CMBfast on a coarser grid and then interpolating 
the multipoles $C_\l$ onto the fine grid.
To prevent the resulting model file from exceeding 9 gigabytes in size,
we also use an adaptive mesh approach, complementing the global grid
with a finer subgrid in the most favored regions of parameter space.
%%% (the regions that affect figures 2 and 3).

\subsection{Separating scalars and tensors}

If we were to run CMBfast in the standard way, computing scalar and tensor
fluctuations simultaneously, we would have to explore an 8-dimensional 
model grid since only $Q$ drops out as an overall normalization factor. 
Instead, we compute 
the scalar fluctuations $\Cl^{scalar}$ and 
the tensor fluctuations $\Cl^{tensor}$
separately, normalize them to both have a quadrupole of unity, and 
compute the combined power spectrum as
\beq{TensorComboEq}
\Cl = Q^2\left[\Cl^{scalar} + r\Cl^{tensor}\right].
\eeq
We therefore only need to compute two 
6-dimensional grids with CMBfast, one over
$(\om,\ob,\tau,h,\Ok,\ns)$ and the other over
$(\om,\ob,\tau,h,\Ok,\nt)$.

In addition, we reduce the dimensionality of our parameter space
to 8 by imposing the consistency relation (Liddle \& Lyth 1992)
\beq{ConsistencyEq}
r = -7 n_t.
\eeq
It holds in all monomial inflation models satisfying 
the slow-roll conditions (the $2^{nd}$ relation
$n_t = n_s-1$ holds only for a small subclass).
We do this merely because it is well-motivated and reduces
error bars -- it does not accelerate our calculations.

\subsection{Separating small and large scales}
\label{HighLowSec}

The multipole moments $\Cl$ for $\l\ll 100$ correspond 
to fluctuations on scales outside the
horizon at recombination. This makes them almost independent of 
the causal microphysics that create the familiar acoustic peaks, 
\ie, independent of $\om$ and $\ob$. We therefore compute 
the power spectrum
for $\l\le 100$ with the fine grid restricted to 
$(\tau,h,\Ok,\ns)$ or $(\tau,h,\Ok,\nt)$, using only an ultra-course 
three-point grid for $\om$ and $\ob$ to pick up weak residual effects aliased down 
from larger $\l$. We then fill in the rest of the $\om$- and $\ob$-values
by interpolation.

For the remaining (high $\l$) part of the power spectrum, 
more radical simplifications can be made.
First of all, the effect of reionization is merely
an overall suppression of $\Cl$ by a constant factor 
$e^{-2\tau}$ on these small scales.
Second, the effect of changing both $\Ok$ and $h$ (and implicitly $\Ol$) 
is merely to shift the power spectrum sideways. This is because the
acoustic oscillations at $z\simgt 1000$ depend only on 
$\om$ and $\ob$, and the geometric projection 
of these fixed length scales onto angular scales $\theta$ in the sky
obeys %(REF)
$\theta(\Om,\Ol,h)\propto 1/\dA(\Om,\Ol,\zlss)$.
%\beq{thetaEq}
%\theta(\Om,\Ol,h) = h/\dA(\Om,\Ol,\zlss).
%\eeq
Here $\dA$ is the luminosity distance 
\beq{dlEq}
\dA% (\Om,\Ol,z) 
= (1+\zlss){S(\kappa I)\over H_0\kappa},\quad
\kappa\equiv \sqrt{|1-\Om-\Ol|},
\eeq
% \beq{zzzEq}
% I(\Om,\Ol,z)
% =\int_0^z A(z')^{-1/2}dz',
% \eeq
% \beq{Aeq}
% A(z)\equiv (1+z)^2(1+\Om z)-z(2+z)\Ol,
% \eeq
\beq{Ieq}
I%(\Om,\Ol,z)
=\int_0^{\zlss} {dz'\over[{(1+z')^2(1+\Om z')-z'(2+z')\Ol]^{1/2}}},
\eeq
where $S(x)\equiv\sinh x$, $x$ and $\sin x$ for open $(\Om+\Ol<1)$,
flat $(\Om+\Ol=1)$ and closed $(\Om+\Ol>1)$ universes, respectively.
We compute $\zlss$, the effective redshift of the last scattering 
surface, using the fit in Appendix E of Hu \& Sugiyama (1996).

$\Om$ and $\Ol$ also modify the
late integrated Sachs-Wolfe effect, but this is important
only for $\l\simlt 30$ (Eisenstein {\etal} 1998). 
The only other effect is a small correction due to gravitational lensing
(Metcalf \& Silk 1998; Stompor \& Efstathiou 1998), which we ignore here because of  
the large error bars on current small-scale data.
To map the model $(\Om^*,\Ol^*,h^*)$ into the model  
$(\Om,\Ol,h)$, one thus shifts its high $\l$ power spectrum
to the right by an $\l$-factor of $\theta(\Om^*,\Ol^*,h^*)/\theta(\Om,\Ol,h)$.
 
We therefore adopt the following procedure.
We compute the $\l\ge 100$ part of the power spectrum for a
3-dimensional grid over
$(\om,\ob,\ns)$ or $(\om,\ob,\nt)$.
We extend this grid to include $h$ and $\Ok$ by shifting it
sideways as described, then merge it with the 
low $\l$ grid by adjusting its normalization to match at $\l=100$.
%%% PUT BACK IF ENOUGH SPACE:
% To avoid ever having to shift to the left (which would require computing 
% additional multipoles), we compute our high $\l$ grid with $h=1.3$
% and $\Ok=0$, the extreme values.  
%%%
In addition to reducing the dimensionality of the grids 
computed with CMBfast, this approach has the advantage that
only flat models need to be run for the high grid, with the
(much slower) computations involving curvature and reionization
only being required up to $\l=100$.

Extensive tests show that these approximations typically reproduce the 
power spectrum to about 5\% accuracy for generic models, 
\ie, substantially better than the current measurement errors. 
As data quality improves, the errors introduced by the above-mentioned
approximation scheme can of course be continuously reduced to zero by
refining the $(\om,\ob)$-grid for low 
$\l$ and shifting the splicing point upwards from $\l=100$.

\smallskip
\smallskip
%\vskip-1.5cm
\centerline{{\vbox{\epsfxsize=8.8cm\epsfbox{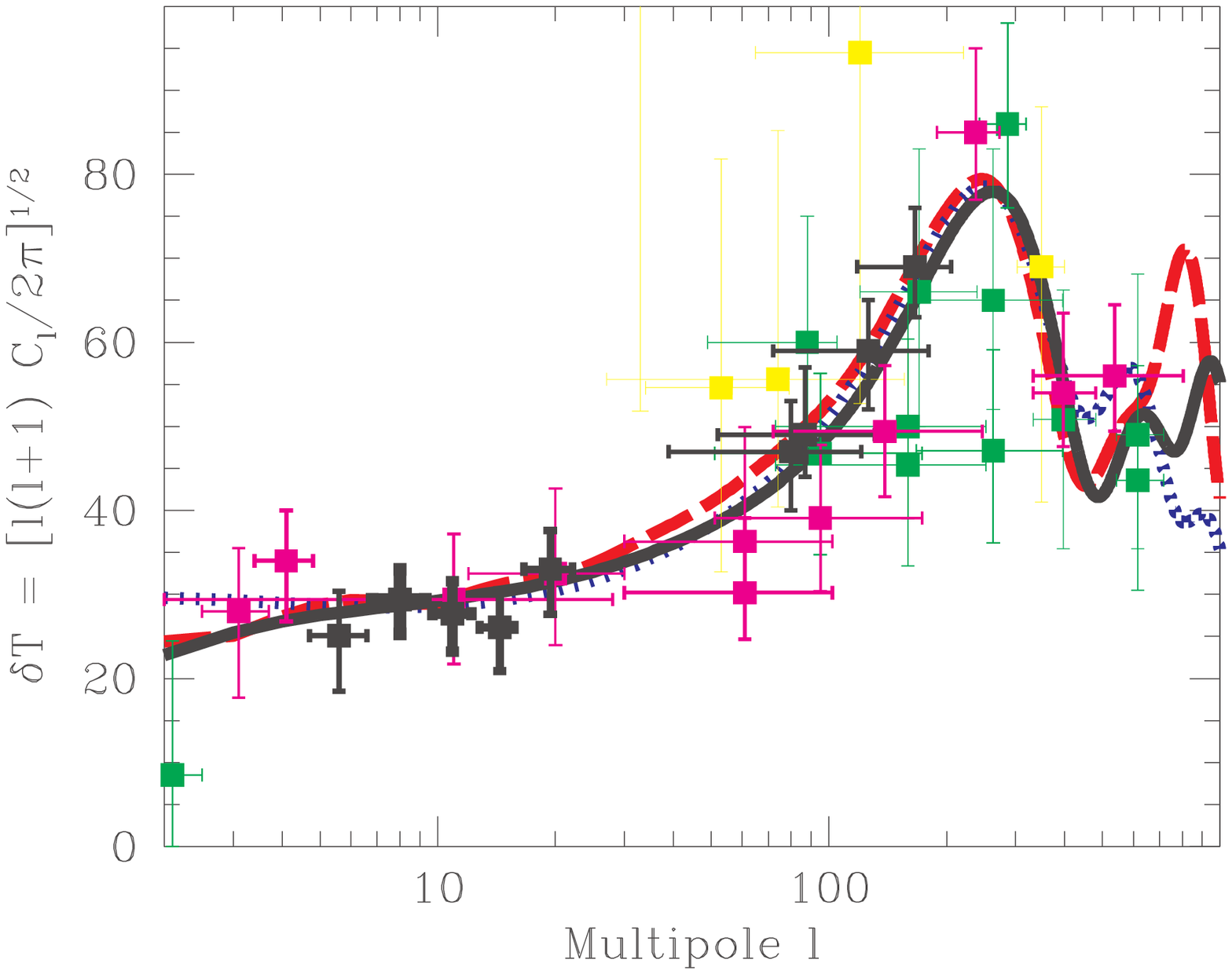}}}}
\vskip-0.3cm
\figcaption{
Three models (see text) are shown 
together with the 37 CMB data points, with the line weight and shading 
emphasizing those with small error bars.
% The best fit (solid line) has 
% $\tau=r=0$, $\Ok=0.3$, $h^2\Om=0.2$, $h^2\Ob=0.025$, $h=0.5$ and $n_s=1.0$.
% The best fit flat model (dashed) has the extreme values
% $\tau=0.5$, $r=0$, $h^2\Om=0.35$, $h^2\Ob=0.040$, $h=0.4$ and $n_s=1.4$,
% yet both it and the 
% ``vanilla'' model (dotted) $\tau=\Ok=r=0$, $n_s=1$,
% $h^2\Om=0.1$, $h^2\Ob=0.02$, $h=0.4$ are seen to differ substantially 
% from the best fit only around the 2nd acoustic peak.
}
%\bigskip

\subsection{Data and likelihoods}

We use the compilation of CMB data and window functions of L98
with the addition of the new QMAP results
(Devlin {\etal} 1998; Herbig {\etal} 1998; de Oliveira-Costa {\etal} 1998),
from which we use the two points combining both flights.
The 37 band powers are shown in Figure 1.
We compute our likelihood function as
$L(\p)\propto e^{-\chi^2/2}$, where 
the $\chi^2$ fit of the data to $C_\l(\p)$
is computed as in L98.
This procedure has a number of deficiencies % , described below.
as we will now describe.
%%%
The probability distributions for the measurements are not Gaussian.
In addition, the error bars for all experiments include a sample variance term
which depends on $\p$, and this dependence is rarely included 
explicitly in quoted measurement results.
A better (offset log-normal) approximation for the band-power likelihood
is given by Bond {\etal} (1998), but
for most experiments, the additional parameter that it requires has 
unfortunately not been computed yet.
%%%
Finally, if the likelihood function $L(\p)$ is a multivariate 
Gaussian, then one can show that marginalizing (integrating) 
over a subset of parameters
is equivalent (up to an irrelevant normalization factor) to maximizing over
them.
% {\bf [PROVE THIS IF SPACE]}
We will follow L98 in doing the latter, since it is both simpler 
and avoids the unpleasant ambiguities of choosing a Bayesian prior
--- alas, with a uniform prior, our 9-dimensional normalization
integral would not even converge. 
As we will see, our $L$ is in fact highly non-Gaussian in some directions, 
which means that our confidence limits must be taken  
with a grain of salt. They also depend on the choice of Bayesian prior,
as described in \S\ref{SingleParSec}.

However, to put these statistical issues in perspective, 
this author feels that an even more pressing challenge will be to test the 
data sets for systematic errors, \eg, by comparing them pairwise
where they overlap in sky coverage and angular resolution 
(Knox {\etal} 1998; Tegmark 1998).

\section{RESULTS AND CONCLUSIONS}

\subsection{Best fit}

The best fit model is shown in Figure 1, and gives 
$\chi^2=22.9$. The probability of obtaining such a low 
$\chi^2$-value with $37-8=29$ effective degrees of freedom 
is about $22\%$, 
so although CMB experimentalists have occasionally been accused of 
underestimating their error bars, we are closer to the opposite situation here.
 
% Although CMB experimentalists are occasionally been accused of 
% underestimating their error bars, we have the opposite problem here:
% the fit is almost too good.
% We have 37 measurements and 8 effective fitting parameters,
% yet the probability for a $\chi^2$-variable with
% 37-8=29 degrees of freedom to give such a low value is
% about $20\%$.
% Possible causes may be systematic reporting of overly cautious error bars
% or, more worrying, a tendency to keep searching for systematic errors 
% until a result near ``the others'' is obtained.
% Hopefully, this problem will disappear as data quality improves in the near
% future --- until then, any elaborate analysis of the data using 
% frequentist statistics appears futile, and will give much weaker constraints
% than a Bayesean likelihood analysis.

It is noteworthy that despite our large parameter space, the
best fit model 
$\tau=r=0$, $\Ok=0.3$, $h^2\Oc=0.2$, $h^2\Ob=0.025$, $h=0.5$ and $n_s=1.0$
(solid line in Figure 1)
is comparatively boring, preferring neither
reionization, gravity waves nor tilt and giving rather conventional
values of $h^2\Ob$ and $h$.
% It has Lambda = -0.2.
Much more exotic models are also allowed, however.
If we restrict the parameter search to flat models ($\Ok=0$), the
best fit is 
$\tau=0.5$, $r=0$, $h^2\Oc=0.35$, $h^2\Ob=0.04$, $h=0.4$ and $n_s=1.4$,
dashed in Figure 1, where the high acoustic peaks that would be caused 
by the strong blue-tilting and the high baryon density are tempered by
very early reionization. If we restrict ourselves to inflationary 
``vanilla'' models with $\tau=\Ok=r=0$ and $n_s=1$, the best fit is
$h^2\Oc=0.1$, $h^2\Ob=0.02$ and $h=0.4$, dotted in Figure 1.
% The figure also shows that these three models only start differing substantially
% at the second acoustic peak, illustrating the importance of more small-scale
% experiments.

\subsection{Single-parameter constraints}
\label{SingleParSec}

Constraints on individual parameters are shown in Figure 2 and Table 1,
interpolating their marginal distributions.
Gravity waves are seen to be generally disfavored, with the maximum-likelihood
value $n_t=0$ corresponding to $r=0$, no gravity waves at all.
The best fitting models all fail to quite match the low 
COBE DMR quadrupole, and tensors merely make this worse by adding additional
large scale power. Reionization is also mildly disfavored, for the same reason
--- increasing $\tau$ and simultaneously increasing $Q$ by a factor $e^\tau$
causes mainly a net rise at small $\l$. 
However, this feature is softer than that of gravity waves, so as 
illustrated in Figure 1, it can be largely offset by increasing 
$\ns$, $\ob$ and $\om$.
The result is that there are no relevant constraints on $\tau$:
not even the extreme case $\tau=0.8$ can be ruled out from our CMB data.

The thin lines show the constraints assuming $\tau=r=0$, as in L98, 
and agree well with the L98 results considering that these did not include
QMAP. However, the heavy lines show that 
including $r$ and $\tau$ substantially weakens these bounds.
Gravity waves and reionization soften the upper limits on $n_s$, $\om$
and $\ob$ 
since they can lower the acoustic peaks given 
COBE-normalization on large scales.

% The curvature bounds are weakened in both ends since they 

% Our results agree well will L98 when considering two differences: our

\bigskip
% \vskip-3.0cm
\centerline{{\vbox{\epsfxsize=8.9cm\epsfbox{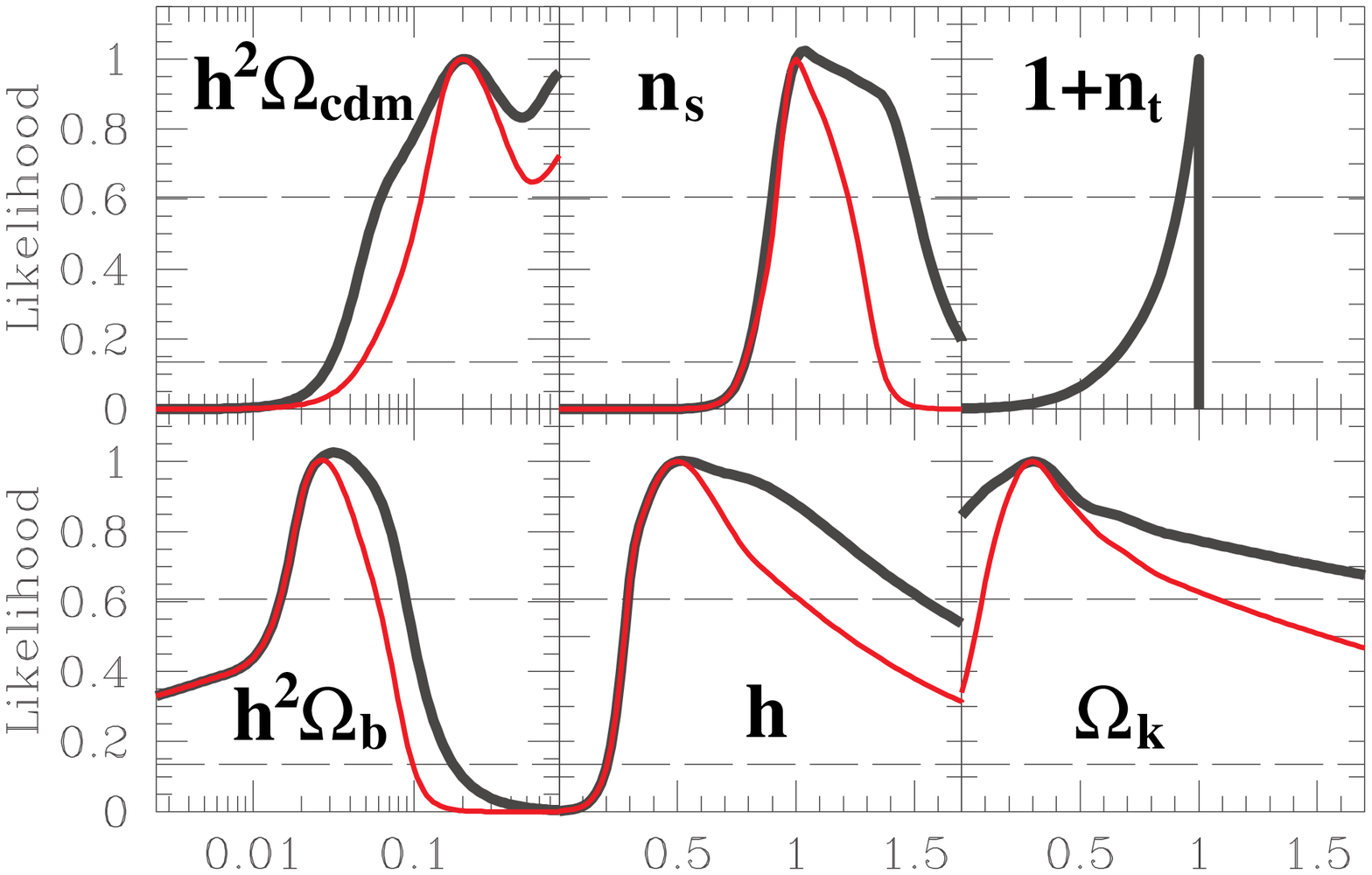}}}}
\figcaption{Heavy lines show likelihoods for individual 
parameters marginalized over all others.
Thin lines show the stronger constraints resulting from assuming 
neither reionization nor gravity waves.
If the likelihood were Gaussian, the 
$68\%$ and $95\%$ confidence limits would lie where the curves 
cross the two dashed lines (see Table 1).
}
%\bigskip
\smallskip

%%%%%%%%%%%%%%%%%%
We have followed L98 in using a 
uniform prior, truncated outside the explored parameter range.
Whereas L98 limited this range to values considered reasonable, 
we have attempted to quantify what CMB alone can say, extending the
range far enough for the likelihood to become small.
Figure 2 shows that this was achieved for all parameters
except $\oc$, where our exclusion of (quite unreasonable)
values $h^2\Oc>0.8$ matters. 
For a full Bayesian analysis, our CMB likelihood function should
be multiplied by the likelihood 
functions from all other relevant astrophysical measurements.
%%%%%%%%%%%%%%%%%%

\subsection{Constraints on the acceleration of the Universe}

The above-mentioned fact that $\Ok$ and $h$ (and implicitly $\Ol$) 
both shift the 
high $\l$ power spectrum sideways make them rather degenerate.
To better understand the constraints on these quantities, 
we therefore plot them in the two-dimensional $\Om-\Ol$ plane 
(Figure 3), where $\Om\equiv\Oc+\Ob$.
Our results for $\tau=r=0$ agree well with those of L98 when considering 
that (a) our analysis includes QMAP and (b) we have plotted 
our 68\% and 95\% confidence contours at $\Delta\chi^2= 2.29$
and $6.18$, respectively, since they 
are two-dimensional, as in Press {\etal} (1992) \S 15.6, 
whereas L98 used $\Delta\chi^2=1$ and $4$.
Unfortunately, CMBfast cannot currently handle closed 
($\Ok<0$) models (White \& Scott 1996). As L98 points out, the likelihood is 
already decreasing as one approaches the diagonal 
$\Ok=0$ line (dotted) from the lower left, so we have simply extended our
likelihood function to $\Ok<0$ by extrapolation.
When dropping the $\tau=r=0$ assumption, 
however, this is no longer true, and the upper right (light grey)
region of the $\Om-\Ol$ plane is no longer excluded. 
% To clarify this situation, it would
% be valuable if CMBfast were upgraded to handle closed models 
% (White \& Scott 1996) as well.

Figure 3 also shows that the constraints at the lower left
are unaffected by reionization and gravity waves. This asymmetry
is easy to understand
physically. This region is ruled out because the first acoustic peak 
is too far to the right,
% to be consistent with the CAT and OVRO data,
whereas the light grey region had the peak too far to the left.
Adding a strong blue-tilt can shift the peak 
slightly to the right, but never to the left. Figure 1 showed that
such a tilted peak could be lowered back to the original height using 
$\tau$ and $r$, making it fit the data, but $\tau$ and $r$ clearly cannot 
raise a (red-tilted) peak.

\smallskip
%\bigskip
\centerline{{\vbox{\epsfxsize=8.9cm\epsfbox{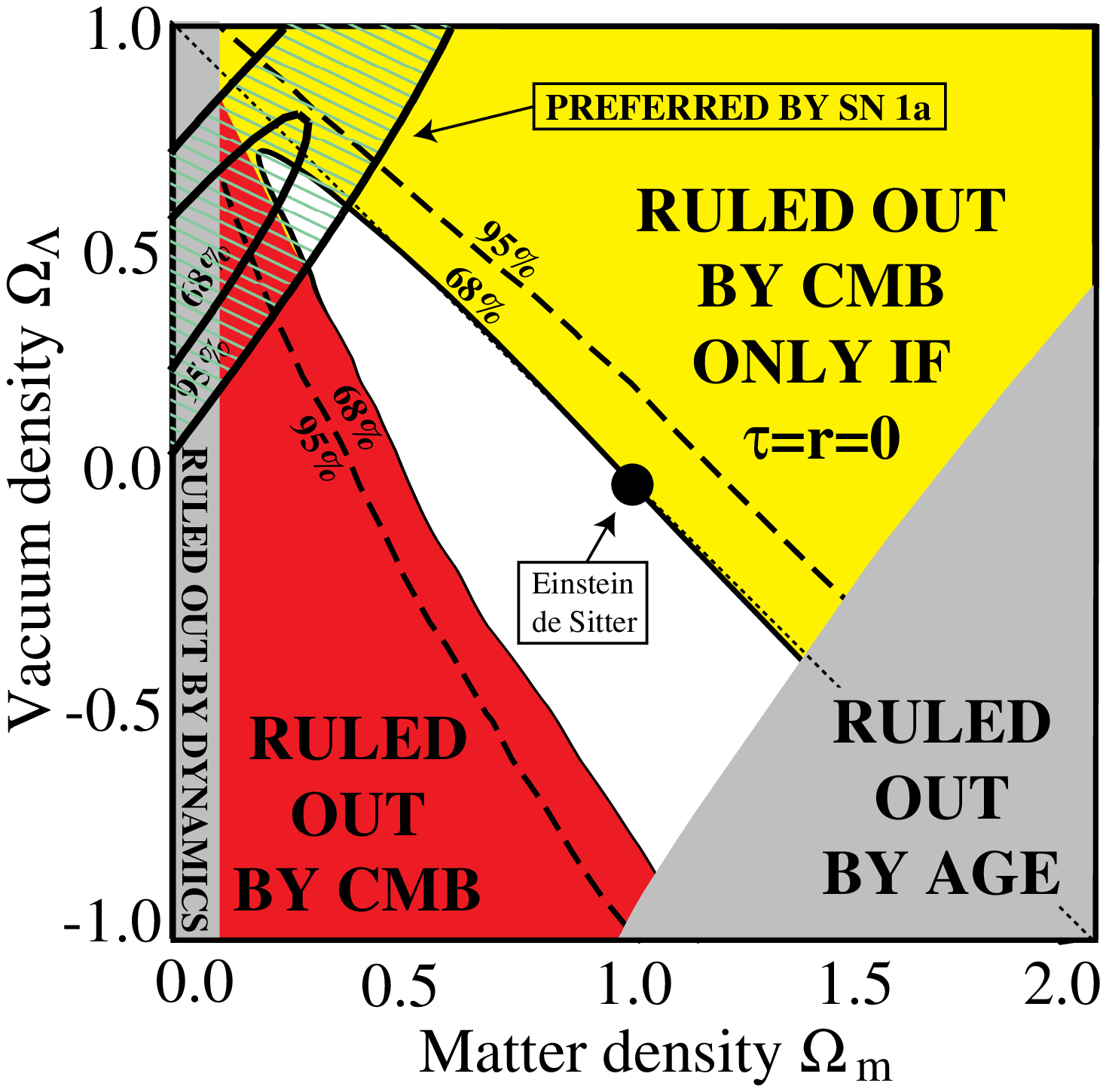}}}}
\vskip-0.2cm
\figcaption{Constraints in the $\Om-\Ol$ plane.
Apart from our CMB constraints, 
$\Om<0.1$ would be inconsistent with the amount of matter 
observed dynamically.
For the age we have simply taken $H_0 t_0>0.6$ as a lower limit.
The region preferred by SN 1a is that computed by White (1998)
from the combined data of the two supernova teams.
In addition, gravitational lensing constrains the upper left
corner. 
}
\bigskip

The recent constraints from SN 1a are highly complementary to our CMB 
constraints. Figure 3 shows the SN 1a constraints computed 
by White (1998) in a joint analysis of the published data from 
the two search teams (Perlmutter {\etal} 1998; Riess {\etal} 1998;
Garnavich {\etal} 1998).
We see that even including $r$ and $\tau$,
the combined CMB and SN 1a data prefer
$\Ol\simgt 0.5$, with a vanishing cosmological 
constant strongly disfavored. The conclusion
that $\Om\simlt 0.5$
does {\it not} survive the inclusion of $r$ and $\tau$,
however, and we cannot rule
out the possibility that the Universe is closed. 

\subsection{Outlook}

In conclusion, we have performed a brute force 8 parameter fit of 
cosmological models to the currently available CMB data
and compared this with SN 1a constraints.
We found that although the inclusion of reionization and gravity waves
weakened many bounds, interesting constraints remain on {\eg} $\Ol$.
Quoted error bars on parameters have grown steadily since the
first COBE results, as more parameters have been included in the analysis.
Since we have now extended our parameter space to essentially the 
full ``minimal cosmological model'',
the error bars might be as large in this {\it Letter} 
as they will ever get. From now on, the rapid improvement in data quality 
will hopefully decrease them faster than they are diluted by 
the addition of further parameters, 
ushering us into the era of precision cosmology.

%  Still more parameters? Comment on neutrinos, running of tilt, helium fraction,
% quintessence eq of state, gdm.
% Why didn't I include $Omega_\nu$?
% If LSND is wrong, we probably don't have enough of them 
% to be relevant.

\def\Qrmsps{Q_{rms,ps}}
\def\lm#1#2#3{$#1^{+#2}_{-#3}$}
\def\na{$-$}
\def\lpeak{\l_{peak}}

\bigskip
\noindent
%\fcolorbox{blue}{yellow}
{\footnotesize
{\bf Table 1} -- Maximum-likelihood values and 68\% confidence limits 
}
\smallskip
{
\begin{tabular}{|l|ccc|ccc|}
\hline
			&\multicolumn{3}{c|}{6 Parameters}
			&\multicolumn{3}{c|}{8 Parameters}\\
Quantity		&Min	&Best	&Max	&Min	&Best	&Max\\
\hline
$h^2\Oc$		&.11	&.20	&$-$	&.063	&.20	&$-$\\	
$h^2\Ob$		&.015	&.027	&.061	&.015	&.032	&.087\\	
% $\tau$		&.011	&.020	&$-$	&.063	&.020	&$-$\\	
$h$			&.29	&.49	&1.01	&.29	&.49	&1.50\\	
$\Ok$			&.09	&.31	&1.05	&$-$	&.31	&$-$\\	
$\ns$			&.92	&.99	&1.22	&.92	&1.04	&1.51\\	
$\nt$			&	&	&	&$-$0.08&0.0	&0.0\\	
$r$			&	&	&	&0.0	&0.0	&.56\\	
\hline		
\end{tabular}
}

\bigskip
The author wishes to thank Charley Lineweaver for 
kindly providing his compilation of CMB data, and 
him and Robert Caldwell, Ang\'elica de Oliveira-Costa, 
Daniel Eisenstein, Wayne Hu, Douglas Scott and Paul Steinhardt
for useful discussions.
Support for this work was provided by
NASA though grant NAG5-6034 and 
Hubble Fellowship HF-01084.01-96A from STScI, operated by AURA, Inc. 
under NASA contract NAS5-26555. 

%%%%%%%%%%%%%%%%%%%%%% REFERENCES: %%%%%%%%%%%%%%%%%%%%%%%%%

%\clearpage
%\end{multicols}


\begin{references}   % 

\rfprep\nn Bartlett J {\etal};1998;astro-ph/9804158 

\rf\nn{de Bernardis} P {\etal};1997;ApJ;480;1

\rf\nnn Bond J R;1995;Phys. Rev. Lett.;74;4369
% preprint astro-ph/9407044
% SIGNAL-TO-NOISE EIGENMODE ANALYSIS OF THE TWO-YEAR COBE MAPS.
% PHYSICAL REVIEW LETTERS, 1995 MAY 29, V74 N22:4369-4372.

\rf\nnn Bond J R, \nn Efstathiou G\multiand \nn Tegmark M;1997;MNRAS;291;L33
% preprint astro-ph/9702100.
% parameters

\rfprep\nnn Bond J R\dualand\nnn Jaffe A H;1998;astro-ph/9809043
% Title: Constraining Large Scale Structure Theories with the Cosmic Background Radiation
% Authors: J. Richard Bond (CITA), Andrew H. Jaffe (Berkeley)
% Comments: 20 pages, LaTeX, 5 figures, 2 tables. To appear in Philosophical Transactions of the Royal Society of London A, 1998.
% "Discussion Meeting on Large Scale Structure in the Universe," Royal Society, London, March 1998. Text and colour figures also available at this ftp URL

\rfprep\nnn Bond J R, \nnn Jaffe A H\multiand\nnn Knox L E;1998;astro-ph/9808264
% Title: Radical Compression of Cosmic Microwave Background Data
% Authors: J. R. Bond (1), A. H. Jaffe (2), L. E. Knox (1) ((1) CITA, U. Toronto, (2) CfPA, Berkeley)
% Comments: 24 Pages, 11 Figures, AASTeX

\rf\nnn Bunn E F\dualand \nn Sugiyama N;1995;ApJ;446;49
% preprint astro-ph/9407069
% Lambda
% COSMOLOGICAL CONSTANT COLD DARK MATTER MODELS AND THE COBE TWO-YEAR SKY MAPS.
% ASTROPHYSICAL JOURNAL, 1995 JUN 10, V446 N1:49-53.

\rf\nnn Bunn E F\dualand\nn White M;1997;ApJ;480;6
% The Four-Year COBE Normalization and Large-Scale Structure
% Authors: Emory F. Bunn (U.C. Berkeley), Martin White (Fermilab)
% Comments: LaTeX, 22 pages. Uses aaspp4.sty. 18 PostScript figures included. Final version,
% accepted to the Astrophysical Journal. Numerous small changse have been made, but all the main
% results are unchanged
% Journal-ref: Astrophys.J. 480 (1997) 6-21

\rfprep\nn Contaldi C, \nn Hindmarsh M\multiand\nn Magueijo J;1998;astro-ph/9809053
%   Title: CMB and density fluctuations from strings plus inflation
%   Authors: Carlo Contaldi, Mark Hindmarsh, Joao Magueijo

\rf\nn{de Oliveira-Costa} A \etal;1998;ApJL;509;L77
% , \nn Devlin M, \nn Herbig T,
% \nnn Miller A D, \nnn Netterfield C B, 
% \nnn Page L A\multiand\nn Tegmark M;1998;ApJL;509;L77
% astro-ph/9808045
% Mapping the Cosmic Microwave Background Anisotropy: Combined Analysis of QMAP Flights
% Authors:
% DE OLIVEIRA-COSTA, ANGELICA; DEVLIN, MARK J.; HERBIG, TOM;
% MILLER, AMBER D.; NETTERFIELD, C. BARTH; PAGE, LYMAN A.; TEGMARK, MAX
% Journal:
% The Astrophysical Journal, Volume 509, Issue 2, pp. L77-L80.
		     
\rf\nn Devlin M \etal;1998;ApJL;509;L69
% , \nn{de Oliveira-Costa} A, \nn Herbig T,
% \nnn Miller A D, \nnn Netterfield C B, 
% \nnn Page L A\multiand\nn Tegmark M;1998;ApJL;509;L69
% astro-ph/9808043
% Mapping the Cosmic Microwave Background Anisotropy: The First Flight of the QMAP
% Experiment
% Authors:
% DEVLIN, MARK J.; DE OLIVEIRA-COSTA, ANGELICA; HERBIG, TOM;
% MILLER, AMBER D.; NETTERFIELD, C. BARTH; PAGE, LYMAN A.; TEGMARK, MAX
% Journal:
% The Astrophysical Journal, Volume 509, Issue 2, pp. L69-L72.


\rfprep\nn Efstathiou G\dualand\nnn Bond J R;1998;astro-ph/9807103
% astro-ph/9807103 [abs, src, ps, other] :
% Title: Cosmic Confusion: Degeneracies among Cosmological Parameters Derived from Measurements of Microwave Background
% Anisotropies
% Authors: G. Efstathiou, J.R. Bond
% Comments: Submitted to MNRAS 25 pages 17 Figures latex mn.sty
       
\rfprep\nnn Eisenstein D J, 
\nn Hu W\multiand Tegmark M;1998;astro-ph/9807130
% parameters2

\rf\nnn Garnavich P M {\etal};1998;ApJ;509;74
% The Astrophysical Journal, Volume 509, Issue 1, pp. 74-79.
% astro-ph/9806396 [abs, src, ps, other] :
% Title: Supernova Limits on the Cosmic Equation of State
% Authors: P. M. Garnavich (1), S. Jha (1), P. Challis (1), A. Clocchiatti (2), A. Diercks (3), A.
% V. Filippenko (4), R. L. Gilliland (5), C. J. Hogan (3), R. P. Kirshner (1), B. Leibundgut (6),
% M.M. Phillips (7), D. Reiss (3), A. G. Riess (4), B. P. Schmidt (8), R. A. Schommer (7), R. C.
% Smith (9), J. Spyromilio (6), C. Stubbs (3), N. B. Suntzeff (7), J. Tonry (10), S. M. Carroll (11)
% ((1) CfA, (2) Univ. Catolica, (3) Univ. Washington, (4) UCB, (5) STScI, (6) ESO, (7) CTIO,
% (8) MSSSO, (9) Univ. Michigan, (10) Univ. Hawaii, (11) UCSB)
% Comments: Accepted for publication in ApJ, 3 figures

\rf\nn Gawiser E\dualand\nn Silk J;1998;Science;280;1405
% astro-ph/9806197 [abs, src, ps, other] :
% Title: Extracting Primordial Density Fluctuations
% Authors: Eric Gawiser, Joseph Silk (U.C. Berkeley)
% Comments: 20 pages including 4 color postscript figures. Full-size figures and data compilation available at this http URL
% Journal-ref: Science 280 (1998) 1405
% CMB + LSS stuff

\rf\nnn Gorski K M {\etal};1994;ApJL;430;L89
% astro-ph/9403067 [abs, ps] :
% On Determining the Spectrum of Primordial Inhomogeneity from the COBE DMR Sky
% Maps: II. Results of Two Year Data Analysis, K.M. Gorski, G. Hinshaw, A.J. Banday, C.L. Bennett, E.L. Wright, A. Kogut, G.F.
% Smoot & P. Lubin. 13 pages, 4 figures included, uuencoded Postscript file. Submitted to ApJ Letters, COBE Preprint #94-08.

% astro-ph/9608054 [abs, src, ps, other] :
% Title: COBE-DMR-Normalized Open CDM Cosmogonies
% Authors: K. M. Gorski (TAC), B. Ratra (KSU), R. Stompor (Oxford), N. Sugiyama (Kyoto), A. J. Banday (MPI fur Astrophysik)
% Comments: 49 pages, uses aaspp4.sty; 2 Postscript files of tables; uuencoded format. Complete paper including text, tables and 24
% figures available at this ftp URL
% APPEARED IN APJS TWO YEARS LATER.

% astro-ph/9512148 [abs, src, ps] :
% Title: CMB Anisotropy in COBE-DMR-Normalized Open CDM Cosmogony
% Authors: Bharat Ratra, Anthony J. Banday, Krzysztof M. Gorski, Naoshi Sugiyama
% Comments: 9 pages including 2 figures, one table: two uuencoded postscript files
% 
% astro-ph/9511087 [abs, src, ps, other] :
% Title: Flat Dark Matter Dominated Models with Hybrid Adiabatic Plus Isocurvature Initial Conditions
% Authors: R.Stompor (CAMK), K.M. Gorski (GSFC), A.J. Banday (GSFC)
% Comments: Two uuencoded compressed Postscript files containing (1) 19 pages manuscript, (2) four figures (tarred together).
% Submitted to The Astrophysical Journal
%        
% astro-ph/9506088 [abs, ps] :
% Title: COBE-DMR-normalization for inflationary flat dark matter models
% Authors: R. Stompor, K.M. Gorski, A.J. Banday
% Comments: uuencoded postscript file (complete text and figures). Accepted for publication in MNRAS
% 
% astro-ph/9502035 [abs, src, ps] :
% Title: COBE-DMR-normalization for Cosmological Constant Dominated Cold Dark Matter Models.
% Authors: Radoslaw Stompor, Krzysztof M. Gorski, Anthony J. Banday.
% Comments: uuencoded file containing 6 postscript files (10 pgs text, 4 figures, 1 table).
% 
% astro-ph/9502034 [abs, src, ps] :
% Title: COBE-DMR-normalized Open Inflation, CDM Cosmogony.
% Authors: Krzysztof M. Gorski, Bharat Ratra, Naoshi Sugiyama, Anthony J. Banday.
% Comments: uuencoded file containing 6 postscript files (11 pgs text, 4 figures, 1 table).
% Journal-ref: Astrophys. J. 444 (1995) L65

% astro-ph/9502033 [abs, src, ps] :
% Title: COBE-DMR-normalization for Cold and Mixed Dark Matter Models Inflationary Cosmogony.
% Authors: Krzysztof M. Gorski, Radoslaw Stompor, Anthony J. Banday.
% Comments: uuencoded file containing 5 postscript files (11 pgs text, 3 figures, 1 table).

\rf\nn Hancock S {\etal};1998;MNRAS;294;L1
% astro-ph/9708254 [abs, src, ps, other] :
% Title: Constraints on cosmological parameters from recent measurements of CMB anisotropy
% Author: S. Hancock (1), G. Rocha (1,3), A.N. Lasenby (1), C.M. Gutierrez (2) ((1) Mullard Radio Astronomy Observatory,
% Cavendish Laboratory (MRAO), (2) Instituto de Astrofisica de Canarias (IAC), (3) Department of Physics, Kansas State University
% (KSU))
% Comments: 7 pages LaTeX, including 6 PostScript figures. Accepted for publication in MNRAS

\rf\nn Herbig T {\etal};1998;ApJL;509;L73
% astro-ph/9808044

% \rf\nn Hu W, \nnn Eisenstein D J\multiand\nn Tegmark M;1998a;
% Phys. Rev. Lett.;xxx;xxxx
% astro-ph/9712057
% neutrinos      

% \rfprep\nn Hu W, \nnn Eisenstein D J, Tegmark M\multiand
% \nn White M;1998b;astro-ph/98xxxxx
% ;ApJ;xxx;xxxx
% gdm   

\rf\nn Hu W\dualand \nn Sugiyama N;1996;ApJ;471;572

\rf\nn Hu W, \nn Sugiyama N\multiand\nn Silk J;1997;Nature;386;37
% preprint astro-ph/9604166.

%\bibitem{Jungman}
\rf\nn Jungman G, \nn Kamionkowski M, \nn Kosowsky A\multiand 
\nnn Spergel D N;1996;Phys. Rev. D;54;1332
% preprint astro-ph/9512139.
% JUNGMAN G; KAMIONKOWSKI M; KOSOWSKY A; SPERGEL DN.
% COSMOLOGICAL-PARAMETER DETERMINATION WITH MICROWAVE BACKGROUND MAPS.
% PHYSICAL REVIEW D, 1996 JUL 15, V54 N2:1332-1344.

\rf\nn Knox L {\etal};1998;Phys. Rev. D;58;083004
% astro-ph/9803272
% \rfprep\nn Knox L, \nnn Bond J R, \nnn Jaffe A H, 
% \nn Segal M\multiand\nn Charbonneau D;1998;astro-ph/9803272
% Comparing Cosmic Microwave Background Datasets
% Journal-ref: Phys.Rev. D58 (1998) 083004

\rfprep\nn Lesgourgues J {\etal};1998;astro-ph/9807019

\rf\nnn Liddle A R\dualand\nnn Lyth D H;1992;Phys. Lett. B;291;391

\rg\nnn Lineweaver C H;1998;ApJL;505;L69;L98
% {astro-ph/9805326 (``L98'')}
% Title: The Cosmic Microwave Background and Observational Convergence in the Omega_m - Omega_lambda Plane
% Author: Charles H. Lineweaver (UNSW,Sydney)
% Comments: This version conforms to the version accepted by the Astrophysical Journal Letters. Minor revisions include references,
% typos and phrasing. 13 pages including 3 Figures
% The Astrophysical Journal, Volume 505, Issue 2, pp. L69-L73.
		     
\rf\nnn Lineweaver C H\dualand\nn Barbosa D;1998a;A\&A;329;799

\rf\nnn Lineweaver C H\dualand\nn Barbosa D;1998b;ApJ;496;624
% astro-ph/9706077 [abs, src, ps, other] :
% What Can Cosmic Microwave Background Observations Already Say About Cosmological Parameters in Open and
% Critical-Density Cold Dark Matter Models?
% Authors: Charles H. Lineweaver (UNSW, Sydney & Strasbourg), Domingos Barbosa (Sussex & Strasbourg)
% Comments: 18 pages with 7 figures, conforms to accepted version in press: Astrophysical Journal, 496, (April 1, 1998). Three of the
% figures have been modified, references updated, typos corrected. This version includes a table of current CMB measurements
       
%\bibitem{Metcalf}
\rf\nnn Metcalf R B\dualand\nn Silk J;1998;ApJ;489;1
% astro-ph/9708059
% Title: Gravitational Magnification of the Cosmic Microwave Background
% Authors: R. Benton Metcalf, Joseph Silk (University of California, Berkeley)
% Comments: to be published in November 1 Astrophysical Journal, 11 pages, two figures
% Astrophysical Journal v.489, p.1

\rf\nn Perlmutter S {\etal};1998;Nature;391;51

\rfbook\nnn Press W H {\etal};1992;Numerical Recipes, 2nd ed.;Cambridge Univ. Press;Cambridge
 
% \rf\nn Ratra B {\etal};1997;ApJ;481;22 

\rfprep\nn Ratra B {\etal};1998;astro-ph/9807298
% Title: ARGO CMB Anisotropy Measurement Constraints on Open and Flat-Lambda CDM Cosmogonies
% Authors: Bharat Ratra, Ken Ganga, Radoslaw Stompor, Naoshi Sugiyama, Paolo de Bernardis, Krzysztof M. Gorski
% Comments: 21 pages of latex. Uses aaspp4.sty. 8 figures included. ApJ in press
             
\rf\nnn Riess A G {\etal};1998;Astron. J.;116;1009
% Observational Evidence from Supernovae for an Accelerating Universe and a Cosmological Constant% astro-ph/9805201
% The Astronomical Journal, Volume 116, Issue 3, pp. 1009-1038.

\rf\nn Seljak U\dualand\nn Zaldarriaga M;1996;ApJ;469;437
% A line of sight approach to Cosmic Microwave Background anisotropies by 

\rf\nnn Smoot G F {\etal};1992;ApJ;396;L1
% \rf Smoot, G.F.; Bennett, C.L.; Kogut, A.; Wright, E.L.; and others.
% Structure in the COBE Differential Microwave Radiometer first-year maps.

\rfprep\nn Stompor R\dualand\nn Efstathiou G;1998;astro-ph/9805294
% astro-ph/9805294 [abs, src, ps, other] :
% Title: Gravitational lensing of CMB anisotropies and cosmological parameters estimation
% Authors: R. Stompor, G. Efstathiou (Institute of Astronomy, Cambridge)
% Comments: 13 pages, 8 figures, uses mn.sty, submitted to MNRAS (5/5/1998)
       
% \rf\nn Tegmark M;1997;Phys. Rev. Lett.;79;3806
% galfisher, preprint astro-ph/9706198

\rfprep\nn Tegmark M;1998;astro-ph/9809001
% comparing

\rf\nn Tegmark M\dualand\nnn Bunn E F;1995;ApJ;455;1
% brute

\rfprep\nn Tegmark M, \nnn Eisenstein D J, \nn Hu W\multiand\nn Kron R;1998;astro-ph/9805117

\rf\nn Webster M {\etal};1998;ApJL;509;L65
% astro-ph/9802109
% Joint Estimation of Cosmological Parameters from
% Cosmic Microwave Background and IRAS Data
% Authors: WEBSTER, A. M.; BRIDLE, S. L.; HOBSON, M. P.; LASENBY, A. N.; LAHAV, O.; ROCHA, G.
% Journal: The Astrophysical Journal, Volume 509, Issue 2, pp. L65-L68.

\rf\nn White M;1998;ApJ;506;495
% astro-ph/9802295
% Complementary Measures of the Mass Density and Cosmological Constant
% Astrophys.J. 506 (1998) 495

\rf\nn White M\dualand\nn Scott D;1996;ApJ;459;415
% astro-ph/9508157 [abs, ps] :
% Title: Why Not Consider Closed Universes?
% Authors: Martin White, Douglas Scott
% Comments: 24 pages, including 13 figures in a uuencoded self-unpacking shell script. Submitted to ApJ
% Journal-ref: Astrophys.J. 459 (1996) 415

%\bibitem{Zalda}
\rf\nn Zaldarriaga M, \nn Spergel D\multiand \nn Seljak U;1997;ApJ;488;1
% Microwave Background Constraints on Cosmological Parameters
% Author(s): M. Zaldarriaga, D. Spergel, U. Seljak

% \rn Gunn \& Petersen ref?

%%% HERE ARE MORE OLD MODEL TESTING/PARAMETER FITTING PAPERS:

%%% THE RATRA ZONE: ONE PAPER PER EXPERIMENT

% astro-ph/9406069 [abs, src, ps, other] :
% CBR ANISOTROPY IN AN OPEN INFLATION, CDM COSMOGONY, by Marc Kamionkowski, Bharat Ratra, David N. Spergel,
% and Naoshi Sugiyama, (12 pages, plain TeX; 3 figures available upon request from the authors), IASSNS-HEP-94/39, PUPT-1470,
% POP-568, CfPA-TH-94-27, UTAP-185.
% 
% astro-ph/9512157 [abs, src, ps] :
% Title: CMB Anisotropy in COBE-DMR-Normalized Flat $\Lambda$ CDM Cosmogony
% Authors: Bharat Ratra, Naoshi Sugiyama
% Comments: 9 pages including 2 figures, one 5 pages table: two uuencoded postscript files
% 
% astro-ph/9512168 [abs, ps] :
% Title: Tentative Appraisal of Compatibility of Small-Scale CMB Anisotropy Detections in the Context of COBE-DMR-Normalized
% Open and Flat $\Lambda$ CDM Cosmogonies
% Authors: Ken Ganga, Bharat Ratra, Naoshi Sugiyama
% Comments: 15 page PostScript file, including 6 included figures. Also available via anonymous ftp from this ftp URL
% 
% astro-ph/9602141 [abs, src, ps] :
% Title: UCSB South Pole 1994 CMB anisotropy measurement constraints on open and flat-Lambda CDM cosmogonies
% Author: Ken Ganga, Bharat Ratra, Josh Gundersen, Naoshi Sugiyama
% Comments: Substantially shortened and rewritten. Accepted by ApJ. PostScript. 49 pages of text + tables. 16 pages of figures
% 
% astro-ph/9702082 [abs, src, ps, other] :
% Title: Large-scale structure in COBE-normalized cold dark matter cosmogonies
% Authors: Shaun Cole (Durham), David H. Weinberg (Ohio), Carlos S. Frenk (Durham), Bharat Ratra (MIT & Kansas)
% Comments: Accepted for publication in MNRAS. (shortened abstract) Also available at this ftp URL      
% 
% astro-ph/9702186 [abs, src, ps, other] :
% Title: Using SuZIE arcminute-scale CMB anisotropy data to probe open and flat-\Lambda CDM cosmogonies
% Authors: K. Ganga, Bharat Ratra, S.E. Church, Naoshi Sugiyama, P.A.R. Ade, W.L. Holzapfel, A.E. Lange, P.D. Mauskopf
% Comments: 17 pages including 4 postscript figures. Latex. Uses aaspp4. ApJ, in press
% 
% astro-ph/9708202 [abs, src, ps, other] :
% Title: MAX 4 and MAX 5 CMB anisotropy measurement constraints on open and flat-Lambda CDM cosmogonies
% Author: Ken Ganga, Bharat Ratra, Mark A. Lim, Naoshi Sugiyama, Stacy T. Tanaka
% Comments: Latex, 37 pages, uses aasms4 style      
% 
% astro-ph/9710270 [abs, src, ps, other] :
% Title: Using White Dish CMB Anisotropy Data to Probe Open and Flat-Lambda CDM Cosmogonies
% Author: B. Ratra, K. Ganga, N. Sugiyama, G.S. Tucker, G.S. Griffin, H.T. Nguyen, J.B. Peterson
% Comments: 17 pages of latex. Uses aasms4.sty. 4 figures included. Submitted to ApJS       
% 
% astro-ph/9807298 [abs, src, ps, other] :
% Title: ARGO CMB Anisotropy Measurement Constraints on Open and Flat-Lambda CDM Cosmogonies
% Authors: Bharat Ratra, Ken Ganga, Radoslaw Stompor, Naoshi Sugiyama, Paolo de Bernardis, Krzysztof M. Gorski
% Comments: 21 pages of latex. Uses aaspp4.sty. 8 figures included. ApJ in press     
% 

% MORE LINEWEAVER:
% astro-ph/9706215 [abs, src, ps, other] :
% Title: Cosmic Microwave Background Anisotropies from Scaling Seeds: Fit to Observational Data
% Authors: R. Durrer, M. Kunz, C. Lineweaver, M. Sakellariadou
% Comments: LaTeX file 4 pages, 4 postscript figs. revised version, to appear in PRL
% Journal-ref: Phys.Rev.Lett. 79 (1997) 5198-5201

\end{references}
\end{document}